\documentclass[aps,prc]{revtex4}
\usepackage{amsmath,amssymb}
\usepackage{graphicx}

\newcommand{\be}{\begin{equation}}
\newcommand{\ee}{\end{equation}}
\newcommand{\ba}{\begin{eqnarray}}
\newcommand{\ea}{\end{eqnarray}}
\newcommand{\ban}{\begin{eqnarray*}}
\newcommand{\ean}{\end{eqnarray*}}

\begin{document}

\title{Sum Rule of Femtoscopic Correlation Function}

\author{Rados\l aw Maj$^1$ and Stanis\l aw Mr\' owczy\' nski$^{1,2}$}

\affiliation{$^1$Institute of Physics, Jan Kochanowski University, ul. \'Swi\c etokrzyska 15, 25-406 Kielce, Poland 
\\
$^2$National Centre for Nuclear Research, ul. Pasteura 7, 02-093 Warsaw, Poland}

\date{November 4, 2019}

\begin{abstract}

A correlation function of two particles with small relative velocities obeys a sum rule - the momentum integral of the function is determined due to the completeness of quantum states of the particles. The original sum rule derived in 1995 suffered from a serious problem: the momentum integral was ultraviolet divergent in physically interesting cases. We resolve the problem by considering the sum rule not of a single correlation function but of a sum or difference of two appropriately chosen correlation functions. The improved sum rule is shown to work well for the exact Coulomb correlation functions. We argue that the sum rule can be used to test an accuracy and range of applicability of correlation functions computed in approximate models. The neutron-proton correlation function is discussed as an example.

\end{abstract}

\maketitle

\section{Introduction}

The correlation functions of two identical or non-identical particles with small relative velocities have been extensively studied in nuclear collisions for bombarding energies from tens of MeV \cite{Boal:yh} to hundreds of GeV \cite{Lisa:2005dd}. These femtoscopic correlation functions provide information about space-time characteristics of particle sources in the collisions. However, when correlations of particles of several species are studied, the functions allow one to infer parameters of inter-particle interaction. In this way the scattering lengths and effective ranges of $p\bar{\Lambda}$, $\bar{p} \Lambda$ and $p\bar{p}$ pairs, which are difficult to access in usual scattering experiments, have been recently measured \cite{Acharya:2019ldv}. 

The correlation function integrated over particle relative momentum was shown \cite{Mrowczynski:1994rn} to satisfy a simple and exact  sum rule which results from the completeness of quantum states. The sum rule relates, for example, the integrated $\pi^\pm \pi^\pm$ correlation function to a pion density in a particle source or the neutron-proton correlation function to a deuteron formation rate. The sum rule qualitatively explains why in spite of attractive inter-particle interaction the correlation can be negative with the correlation function smaller than unity. It happens, for example, in case of $p\bar{p}$ \cite{Acharya:2019ldv} and neutron-proton \cite{Mrowczynski:1994rn} because of possible annihilation or deuteron formation, respectively. More generally, an attractive inter-particle interaction can lead to negative correlations if the interaction is absorptive. 

The sum rule requires to integrate the correlation function up to infinite momentum, but one expects that the integral saturates at a sufficiently large momentum. In case of free identical particles, when the correlation occurs due quantum statistics, the sum rule was shown \cite{Maj:2004tb} to be satisfied if the integration extends up to the momentum $q_{\rm max}$ equal a few times the inverse radius of particle source.  

When the sum rule was applied to correlation functions of interacting particles \cite{Maj:2004tb} an unexpected difficulty was discovered: the integral over relative momentum was badly divergent because the correlation functions tended to unity with growing momentum not sufficiently fast. Therefore, the original derivation of the sum rule \cite{Mrowczynski:1994rn}, which implicitly assumed an existence of the momentum integral, was mathematically flawed. 

The aim of this paper is to present a resolution of the problem which relies on the following observation. While the momentum integral of a single correlation function is divergent, the integral of a difference of two correlation functions of the same asymptotic is convergent. We derive here the sum rule not of a single correlation function but of a difference or sum of two appropriately chosen correlation functions. In particular, we consider the difference of the correlation function of identical charged particles and of non-identical ones. Another pair of interest are the correlation functions of neutron and proton in spin triplet and singlet states. The sum rule of the difference or sum of two correlation functions carries similar physical information as the original one. Since the sum rule is exact it can be used to test a range of applicability and accuracy of models used to compute the correlations functions. 

We start our considerations in Sec.~\ref{sec-prelimi} introducing main concepts of femtoscopy to be used further on. In Sec.~\ref{sec-sum-rule} we derive the original sum rule just to set the stage for the discussion. The modified sum rule is derived in Sec.~\ref{sec-sum-rule-impr} and in subsequent section \ref{sec-test} we show that the improved sum rule properly works when applied to $\pi^\pm \pi^\pm$ and $\pi^+ \pi^-$ correlation functions which are known in an exact form as the Coulomb problem is exactly solvable.  Finally,  in Sec.~\ref{sec-np} we use the sum rule to discuss the neutron-proton correlation functions which are computed within an approximate model. The sum rule reveals a range of applicability of the model and its accuracy. We close the paper with the summary of our main results and conclusions.

\section{Preliminaries}
\label{sec-prelimi}

The correlation function of two particles $a$ and $b$ with momenta ${\bf p}_a$, ${\bf p}_b$ is denoted as $\mathcal{R}({\bf p}_a, {\bf p}_b)$ and it is defined in the following way 
\be
\frac{dP_{ab}}{d^3p_a  d^3p_b} = \mathcal{R}({\bf p}_a , {\bf p}_b) \, 
\frac{dP_a}{d^3p_a}  \frac{dP_b}{d^3p_b} ,
\ee
where $\frac{dP_a}{d^3p_a}$,  $\frac{dP_b}{d^3p_b}$ and $\frac{dP_{ab}}{d^3p_a  d^3p_b}$ are probability densities to observe a particle $a$, a particle $b$ and a pair $ab$ in a collision final state. 

If the inter-particle correlations are due to quantum statistics and/or final state interactions, the correlation function is known to be an overlap integral of the probability distribution of space-time emission points of the particles and the modulus square of the wave function of two particles in the final state \cite{Koonin:1977fh,Lednicky:1981su}. To avoid complications, which are irrelevant for our considerations, we write down the correlation function in the simplest form as 
\be
\label{RA}
\mathcal{R}({\bf p}_a, {\bf p}_b)
=\int d^3r_a \, d^3r_b \,
D({\bf r}_a) \, D({\bf r}_b) \, |\psi ({\bf r}_a,{\bf r}_b)|^2,
\ee
where  the source function $D({\bf r})$ is the probability distribution of emission points of a particle $a$ or $b$ and $\psi ({\bf r}_a,  {\bf r}_b)$ is the wave function of the $ab$ pair. Further on, the source function $D({\bf r})$ is assumed to be the isotropic Gaussian distribution  
\be
\label{Gauss-source-1}
D({\bf r}) = {1 \over (2 \pi r_0^2)^{3/2}}\,
{\rm exp} \Big(-{{\bf r}^2 \over 2r_0^2}\Big) ,
\ee
which gives the mean radius squared of a source equal  $\langle {\bf r}^2 \rangle = 3 r_0^2$.

Let us discuss the formula (\ref{RA}). We first note that it is of non-relativistic form even though the particles of interest are typically relativistic. A relativistic description of strongly interacting particles faces some difficulties particularly severe when bound states are involved. The correlation function, however, significantly differs from unity only for small relative velocities. Therefore, the relative motion can be treated as non-relativistic and the corresponding wave function is a solution of the Schr\"odinger equation. So, we consider the inter-particle correlations in the center-of-mass frame of the pair and consequently, the  source function, which is usually defined in the source rest frame, needs to be transformed to the center-of-mass frame of the pair as discussed in great detail in \cite{Maj:2009ue}. To simplify our analysis we ignore the need to transform the source function and we use the form (\ref{Gauss-source-1}) in any reference frame. 

We also note that the formula (\ref{RA}) is written as for the instantaneous emission of the two particles as the source function is time independent. However, a time duration of the emission process can be easily incorporated following \cite{Koonin:1977fh}. After  taking the time integral the temporal extension of the particle source contributes to the spatial extension. Specifically, if one uses an isotropic source function, the time duration $\tau$ simply enlarges the effective radius of the source from $r_0$ to $\sqrt{r_0^2 + v^2\tau^2}$ where $v$ is the velocity of the particle pair relative to the source. Therefore, a purely spatial source should be treated as an effective source which takes into account a temporal duration of the emission process through an extended spatial size. 

As discussed in detail in \cite{Shapoval:2013bga}, a realistic source function is spatially anisotropic and time dependent. However, we use the simple Gaussian form (\ref{Gauss-source-1}) because our objective is to merely test the sum rule. An analysis of experimental data would obviously require a more realistic parameterization. Nevertheless, it should be clearly stated that the sum rule \cite{Mrowczynski:1994rn} can be applied to any source function, to anisotropic and time dependent as well.  We return to this point when the sum rule is discussed in the next section.

To simplify the formula (\ref{RA}) one eliminates a center-of-mass motion of the particle pair. For this purpose we introduce the center-of-mass variables 
\be
\label{CM-2-variables}
{\bf R} \equiv \frac{m_a {\bf r}_a + m_b{\bf r}_b }{M} ,
~~~~~~
{\bf r} \equiv  {\bf r}_a - {\bf r}_b ,
\ee
where $M \equiv m_a + m_b$. We write down the wave function as $\psi ({\bf r}_a,{\bf r}_b) = e^{i {\bf R} {\bf P}} \phi_{\bf q} ({\bf r})$ with ${\bf P}$ and ${\bf q}$ being the momentum of the center of mass and the momentum in the center-of-mass frame of the pair. The correlation function (\ref{RA}) is then found to be
\be
\label{fun-corr-relative}
\mathcal{R}({\bf q}) = \int d^3 r \, D_r ({\bf r}) |\phi_{\bf q}({\bf r})|^2 ,
\ee
where the `relative' source is 
\be
\label{D-r-source-def}
D_r({\bf r}) \equiv \int d^3 R \, D \Big ({\bf R} + \frac{m_a}{M}{\bf r} \Big) 
\, D \Big ({\bf R}-\frac{m_b}{M} {\bf r} \Big ) .
\ee
One checks that if the single particle source is normalized, that is $\int d^3r D({\bf r}) = 1$, the `relative' source is normalized as well. We also note that when the single-particle source function is of the Gaussian form (\ref{Gauss-source-1}), the `relative' source is also Gaussian and it equals
\be 
\label{Gauss-source-r}
D_r({\bf r}) = {1 \over (4 \pi r_0^2)^{3/2}}\,
{\rm exp} \Big(-{{\bf r}^2 \over 4r_0^2}\Big) .
\ee

\section{Sum Rule}
\label{sec-sum-rule}

The sum rule, which is the main theme of the paper, was first derived in \cite{Mrowczynski:1994rn}. Let us repeat the original derivation as a starting point of our analysis. 

One considers a correlation function integrated over a relative momentum of the two particles. Since ${\cal R}({\bf q}) \rightarrow 1$ when ${\bf q} \rightarrow \infty$, one rather considers the integral of ${\cal R}({\bf q}) - 1$. Using Eq.~(\ref{fun-corr-relative}) and taking into account that the source function is normalized to unity, one finds, after changing the order of the ${\bf r}-$ and  ${\bf q}-$integration, the expression 
\be 
\label{sum1}
\int {d^3 q \over (2\pi )^3} \, \Big( {\cal R} ({\bf q}) - 1 \Big)
= \int d^3r \, D_r ({\bf r}) \, \int {d^3 q \over (2\pi )^3} \,
\Big( \vert \phi_{\bf q}({\bf r}) \vert ^2 - 1 \Big) .
\ee
We note that changing the order of integrations implicitly assumes that both integrals exist. As already mentioned in the introduction, the assumption appears to be violated in physically interesting situations. 

The integral over ${\bf q}$ in the r.h.s. of Eq.~(\ref{sum1}) is determined by the quantum-mechanical completeness condition. Indeed, the wave functions satisfy the well-known closure relation
\be 
\label{complete1}
\int {d^3 q \over (2\pi )^3} \;
\phi_{\bf q}({\bf r}) \phi^*_{\bf q}({\bf r}') +
\sum_{\alpha} \phi_{\alpha}({\bf r}) \phi^*_{\alpha}({\bf r}')
= \delta^{(3)}({\bf r} - {\bf r}') \pm \delta^{(3)}({\bf r} + {\bf r}') ,  
\ee 
where $\phi_{\alpha}$ represents a possible bound state of the two particles of interest. When the particles are not identical the second term in the r.h.s. of Eq.~(\ref{complete1}) is not present. This term guarantees the right symmetry for both sides of the equation for the case of identical particles. The upper sign is for bosons while the lower one for fermions. The wave function of identical bosons (fermions) $\phi_{\bf q}({\bf r})$ is (anti-)symmetric when ${\bf r} \rightarrow -{\bf r}$, and the r.h.s of Eq.~(\ref{complete1}) is indeed (anti-)symmetric when ${\bf r} \rightarrow -{\bf r}$ or ${\bf r}' \rightarrow -{\bf r}'$. If the particles of interest carry spin, the summation over spin degrees of freedom in the l.h.s. of Eq.~(\ref{complete1}) is implied.  

Even if one considers particles $a$ and $b$ with small relative momentum, the completeness condition can be much more complicated than that expressed in Eq.~(\ref{complete1}). For example, in a $K^- p$ scattering there is an open inelastic channel $K^-p \rightarrow \pi^0\Lambda$ at vanishing momentum. Therefore, the $\pi^0\Lambda$ states should be included in the complete set of states of the $K^- p$ pair. In case of proton-antiproton pair the complete set of states includes multi-pion states because of a possible annihilation. We do not consider such situations and we stay with the simple completeness condition (\ref{complete1}). 

When the integral representation of $\delta^{(3)}({\bf r} - {\bf r}') $ is used, the relation (\ref{complete1}) can be rewritten as
\be
\int {d^3 q \over (2\pi )^3} \, \Big(
\phi_{\bf q}({\bf r}) \phi^*_{\bf q}({\bf r}') 
- e^{i{\bf q}({\bf r}-{\bf r}')} \Big) 
+ \sum_{\alpha} \phi_{\alpha}({\bf r}) \phi^*_{\alpha}({\bf r}') 
= \pm \delta^{(3)}({\bf r} + {\bf r}') .
\ee
Now, we take the limit ${\bf r}' \rightarrow {\bf r}$ and get the relation 
\be 
\label{complete2} 
\int {d^3 q \over (2\pi )^3} \,
\Big( \vert \phi_{\bf q}({\bf r}) \vert ^2 - 1 \Big)
=  \pm \; \delta^{(3)}(2 {\bf r})
- \sum_{\alpha} \vert \phi_{\alpha}({\bf r}) \vert ^2 . 
\ee

When the formula (\ref{complete2}) is substituted into Eq.~(\ref{sum1}), one finds the desired sum rule
\be 
\label{sum}
\int d^3 q \, \Big( {\cal R} ({\bf q}) - 1 \Big)
= \pm \pi^3 \, D_r (0) - \sum_{\alpha} A_{\alpha} , 
\ee
where $A_{\alpha}$ is the formation rate of a bound state ${\alpha}$ 
\be 
\label{form-rate}
A_{\alpha} =  (2\pi)^3 \int d^3r \, D_r({\bf r}) \vert \phi_{\alpha}({\bf r}) \vert ^2 .
\ee
The formation rate relates a probability to observe a bound state ${\alpha}$ of particles $a$ and $b$ with total momentum ${\bf P}$ to the probability to observe the pair of particles $a$ and $b$ with the momenta ${\bf p}_a$ and ${\bf p}_b$ in the following way
\be
\frac{dP_\alpha}{d^3P} = A_\alpha \frac{dP_a}{d^3p_a}\frac{dP_b}{d^3p_b} .
\ee
The momenta are assumed to obey the conditions ${\bf P} = {\bf p}_a + {\bf p}_b$ and ${\bf q}=0$, which physically mean that the particles $a$ and $b$ form the bound state $\alpha$ when their relative velocity vanishes, or rather when $|{\bf P}| \gg |{\bf q}|$. 

The completeness condition is valid for any inter-particle interaction. It is also valid when the pair of particles interact with the time-independent external field, {\it e.g.} the Coulomb field generated by the particle source. The sum rule (\ref{sum}) holds under very general conditions as long as the basic formula (\ref{fun-corr-relative}) is justified, in particular as long as the source function $D_r ({\bf r})$ is ${\bf q}-$independent and spin independent. The validity of these assumptions can be tested  only within a microscopic model of nucleus-nucleus collision which properly describes quantum correlations of particles and bound state formations. 

As already noted, the sum rule as derived in \cite{Mrowczynski:1994rn} is applicable to any source function which can be time dependent and anisotropic. If the source function depends on both space and time position as $D({\bf r},t)$, the expression $D_r (0)$ in the r.h.s. of Eq.~(\ref{sum}) should be replaced by $\int dt \, D_r ({\bf v}t,t)$ where ${\bf v}$ is the velocity of the pair relative to the source. The formation rate (\ref{form-rate}) should be appropriately modified as well \cite{Mrowczynski:1994rn}.

Although the sum rule (\ref{sum}) requires the integration up to infinite momentum, one expects that the integral in Eq.~(\ref{sum}) saturates at a sufficiently large upper limit $q_{\rm max}$ that is the sum rule holds for the function 
\be 
\label{smax}
S(q_{\rm max}) \equiv 4\pi \int_0^{q_{\rm max}} dq \, q^2 \big({\cal R}(q) - 1 \big) ,
\ee
where the source function is assumed to be spherically symmetric. In such a case the correlation function depends only on $q \equiv |{\bf q}|$ and the angular integral is trivial. 
 
The correlation function of non-interacting particles differs from unity only when the particles are identical. The wave function, which enters the correlation function (\ref{fun-corr-relative}), is the plane wave (anti-) symmetrized for bosons (fermions). For the Gaussian source (\ref{Gauss-source-r}), the correlation function equals
$$
{\cal R}(q) = 1 \pm e^{- 4 r_0^2q^2} ,
$$
where the plus sign is for bosons and the minus for fermions, and the function (\ref{smax}) is \cite{Maj:2004tb} 
$$
S(q_{\rm max}) = \pm  \bigg( \frac{\sqrt{\pi}}{2r_0} \bigg)^3
\bigg[ 1 - \frac{4r_0 q_{\rm max}}{\sqrt{\pi}} \, 
e^{- 4 r_0^2 \, q^2_{\rm max}}  
\Big( 1 + {\cal O}\Big(\frac{1}{r_0^2 q_{\rm max}^2} \Big) \bigg] .
$$
Since $ \pi^3 \, D_r (0) = (\sqrt{\pi}/2r_0)^3$, one sees that the sum rule is satisfied for $r_0 q_{\rm max} \gg 1$.

Our attempts to apply the sum rule (\ref{sum}) to physically interesting cases of interacting particles badly failed as described in detail in \cite{Maj:2004tb}. Eq.~(\ref{smax}) clearly shows what was the problem. The momentum integral of the sum rule  (\ref{sum}) exists if the correlation function tends to unity faster than $q^{-3}$ as $q$ grows. Otherwise the integral diverges and the interchange of the ${\bf q}-$ and ${\bf r}-$integration, when the sum rule is derived, is mathematically  incorrect. In the subsequent section we show how to circumvent the difficulty.

\section{Improved sum rule}
\label{sec-sum-rule-impr}

Let us consider not one but two correlation functions ${\cal R} ({\bf q})$ and $\widetilde{\cal R} ({\bf q})$. The latter function is called a {\it regulator}. We choose the functions in such a way that the difference ${\cal R} ({\bf q}) - \widetilde{\cal R} ({\bf q})$ or the sum ${\cal R} ({\bf q}) + \widetilde{\cal R} ({\bf q})$ tends to 0 or 2, respectively, faster than $q^{-3}$ as $q$ grows. Then, the momentum integral analogous to that from Eq.~(\ref{sum}) is convergent. We discus the difference of the correlation functions. The case of the sum is fully analogous. 

Using Eq.~(\ref{fun-corr-relative}) the momentum integral of the difference of the correlation functions is
\be 
\label{sum101}
\int {d^3 q \over (2\pi )^3} \, \bigg[\Big( {\cal R} ({\bf q}) - 1 \Big) - \Big( \widetilde{\cal R} ({\bf q}) - 1 \Big) \bigg]
= \int d^3r \bigg[D_r ({\bf r}) \int {d^3 q \over (2\pi )^3} \,
\Big( \vert \phi_{\bf q}({\bf r}) \vert ^2 - 1 \Big) -
\widetilde{D}_r ({\bf r}) \int {d^3 q \over (2\pi )^3} \,
\Big( \vert \widetilde{\phi}_{\bf q}({\bf r}) \vert ^2 - 1 \Big) \bigg].
\ee
The momentum and position integrals are assumed to converge and thus we can change the order of integration over ${\bf q}$ and over ${\bf r}$. In principle, the source functions $D_r ({\bf r})$ and $\widetilde{D}_r ({\bf r})$ can be different from each other but in practical applications discussed in Secs.~\ref{sec-test} and \ref{sec-np} we consider only the case when $D_r ({\bf r}) = \widetilde{D}_r ({\bf r})$.

Now, we are going to substitute the completeness relation (\ref{complete2}) into Eq.~(\ref{sum101}) and we assume that the correlation function $\widetilde{\cal R} ({\bf q})$ describes a pair of nonidentical particles which do not form any bound state. Then, 
\be
\label{complete-simple}
\int {d^3 q \over (2\pi )^3} \,
\Big( \vert \widetilde{\phi}_{\bf q}({\bf r}) \vert ^2 - 1 \Big)
= 0 ,
\ee
and Eq.~(\ref{sum101}) provides the desired sum rule
\be 
\label{sum-improved}
\int d^3 q \, \Big( {\cal R} ({\bf q}) - \widetilde{\cal R} ({\bf q}) \Big)
= \pm \pi^3 \, D_r (0) - \sum_{\alpha} A_{\alpha} . 
\ee

Further on we will discuss the sum rule (\ref{sum-improved}) but if the completeness relation of the wave functions $\widetilde{\phi}_{\bf q}({\bf r})$ is not given by Eq.~(\ref{complete-simple}) but it is of more general form (\ref{complete2}), one has to subtract the expression  $\pm \pi^3 \, \widetilde{D}_r (0) - \sum_{\alpha} \widetilde{A}_{\alpha}$ from the r.h.s. of Eq.~(\ref{sum-improved}). If one considers not the difference  but the sum of the correlation functions, the integrand in the l.h.s. of Eq.~(\ref{sum-improved}) is ${\cal R} ({\bf q}) + \widetilde{\cal R} ({\bf q}) - 2$ and the expression  $\pm \pi^3 \, \widetilde{D}_r (0) - \sum_{\alpha} \widetilde{A}_{\alpha}$ should be added to the r.h.s. of Eq.~(\ref{sum-improved}). 

In the subsequent sections we will discuss some examples of the correlation functions showing that  the function  
\be 
\label{smax-improved}
{\cal S}(q_{\rm max}) \equiv 4\pi \int_0^{q_{\rm max}} dq \, q^2 \big({\cal R}(q) - \widetilde{\cal R} (q) \big) 
\ee
satisfies the sum rule (\ref{sum-improved}) for a sufficiently big upper limit $q_{\rm max}$.

\begin{figure}[t]
\begin{minipage}{86mm}
\centering
\includegraphics[scale=0.65]{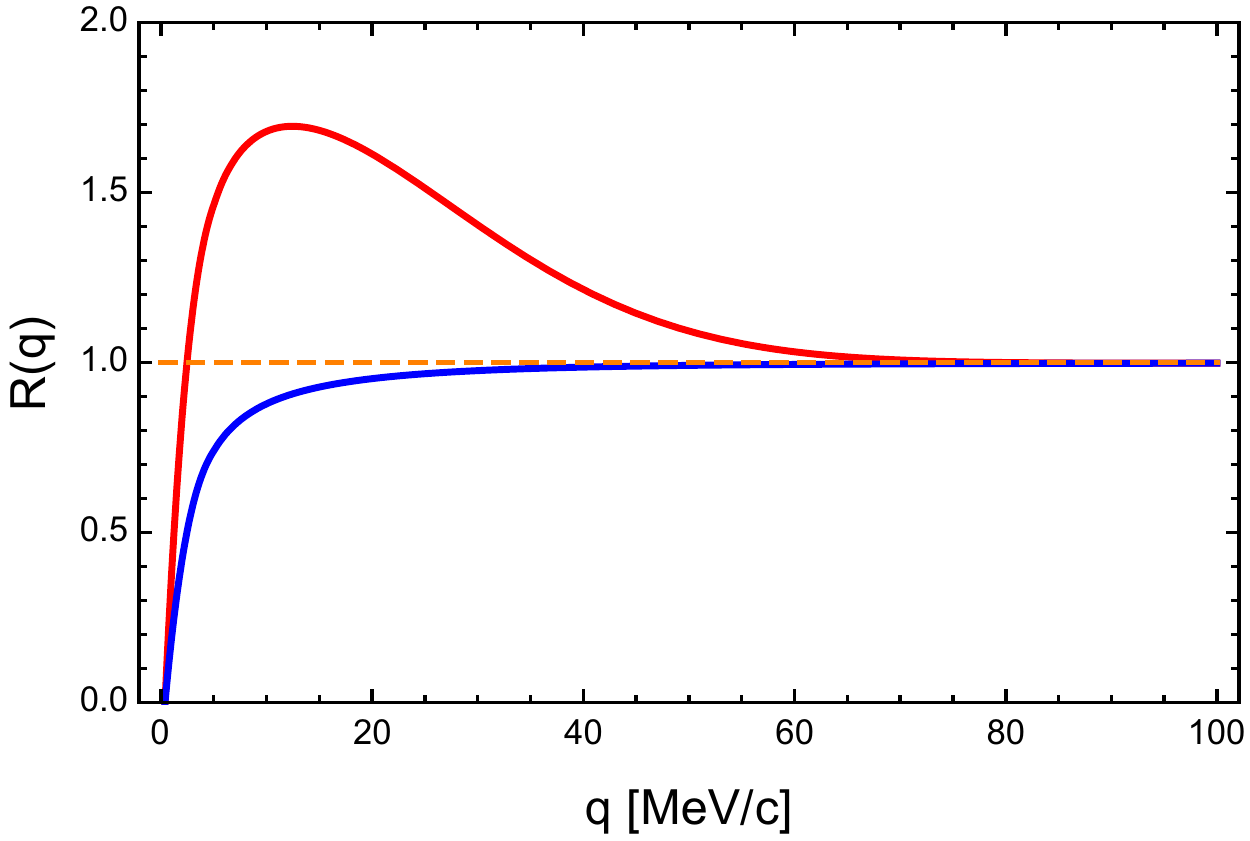}
\caption{The correlation functions of charged identical (red) and non-identical (blue) pions for $r_0 = 3$~fm.}
\label{fig-pion-corr}
\end{minipage}
\hspace{3mm}
\begin{minipage}{86mm}
\centering
\includegraphics[scale=0.68]{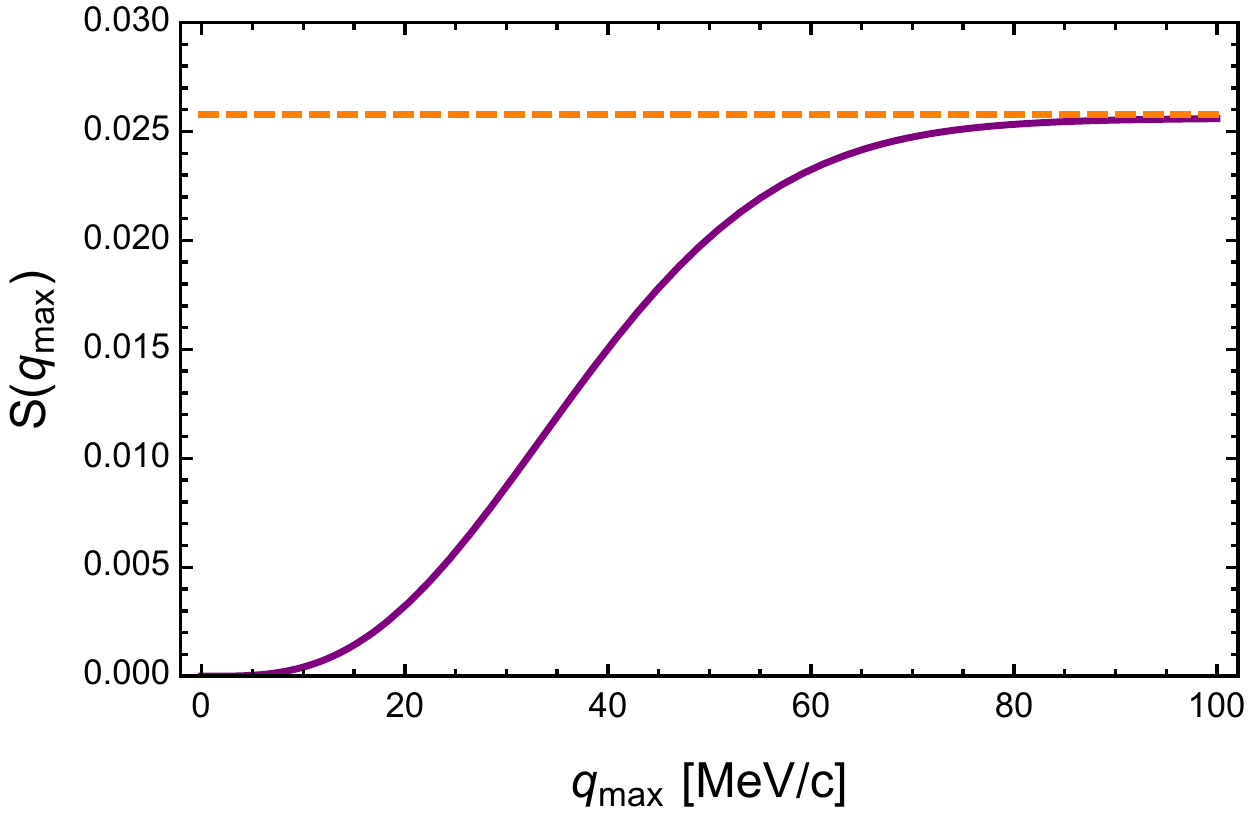}
\caption{The function ${\cal S}(q_{\rm max})$ of charged pions (solid) and the value of $\pi^3 \, D_r (0)$ (dashed) for $r_0 = 3$~fm.}
\label{fig-smax-pions}
\end{minipage}
\hspace{2mm}
\end{figure}

\section{Tests of the sum rule}
\label{sec-test}

As well known, the Coulomb problem is exactly solvable within the non-relativistic quantum mechanics \cite{Schiff68}. Therefore, the correlation functions of charged particles, which experience a Coulomb attraction or repulsion, can be calculated exactly if no other inter-particle interaction is present, see {\it e.g.} \cite{Maj:2004tb}. For definitiveness we consider correlations of charge pions ignoring their strong interaction. Having the exact correlation functions we can test whether the improved sum rule (\ref{sum-improved}) works properly. 

\subsection{Coulomb repulsive interaction}

The correlation function of the same sign pions is denoted as ${\cal R} (q)$ and the regulator $\widetilde{\cal R} (q)$ corresponds to the same charge pions which, however, are not treated as identical that is their wave function is not symmetrized. According to the sum rule  (\ref{sum-improved}), the expected relation is 
\be 
\label{sum-improved-pions}
4\pi \int_0^\infty dq \,q^2  \Big( {\cal R} (q) - \widetilde{\cal R} (q) \Big)
=  \pi^3 \, D_r (0) . 
\ee

A computation of the correlation function of identical and non-identical pions is described in detail in \cite{Maj:2004tb}. In Fig.~\ref{fig-pion-corr} we show the two correlation functions for the Gaussian source function (\ref{Gauss-source-r}) with $r_0 = 3$~fm.  Fig.~\ref{fig-smax-pions} presents the corresponding function ${\cal S}(q_{\rm max})$ given by Eq.~(\ref{smax-improved}) and the value of $\pi^3 \, D_r (0)$. As one observes, ${\cal S}(q_{\rm max})$ becomes indistinguishable from $\pi^3 \, D_r (0)$ for $q_{\rm max} r_0 \ge 1.5$. The value of  $\pi^3 \, D_r (0)$ as a function of $r_0$ is shown Fig.~\ref{fig-D0}. The quantity is computed directly from the source function definition (\ref{Gauss-source-r}) and from the sum rule (\ref{sum-improved-pions}). As expected, the sum rule  (\ref{sum-improved-pions}) is seen to be fulfilled.

\subsection{Coulomb attractive interaction}

The correlation function of opposite charge pions is of special interest because the pionic atoms can occur in this case. As a regulator we choose the correlation function of $\pi^\pm \pi^\pm$ with the wave function which is not symmetrized. The $\pi^+ \pi^-$ correlation function goes to unity from above when $q$ grows while the $\pi^+ \pi^+$ correlation function goes to unity from below.  Therefore, as discussed in Sec.~\ref{sec-sum-rule-impr}, we expect the following sum rule
\be 
\label{sum-improved-pions-att}
4\pi \int_0^\infty dq \,q^2  \Big( {\cal R} (q) + \widetilde{\cal R} (q) - 2 \Big)
= -  \sum_{n=1}^\infty \sum_{l=0}^{n-1} \sum_{m= -l}^{l} A_{nlm} , 
\ee
where $A_{nlm}$ is the formation rate of an atomic state of quantum numbers $(n,m,l)$.

We have not been able to convincingly show by means of numerical analysis that the  momentum integral in Eq.~(\ref{sum-improved-pions-att}) is convergent and that the sum rule (\ref{sum-improved-pions-att}) is satisfied for a finite size source. We have considered instead the case of point-like source when $D_r ({\bf r}) = \delta^{(3)}({\bf r})$. Then, the correlation function of two opposite and same charge particles equals the Gamov factor that is
\be 
\label{Gamov}
G^{\pm}(q) = \pm {2 \pi \over a_B q} \,
{1 \over {\rm exp}\big(\pm {2 \pi \over a_B q}\big) - 1} ,
\ee
where the plus (minus) sign is for the repelling (attracting) particles, $a_B = 1/\mu \alpha$ is the Bohr radius of the two particles, $\mu$ is their reduced mass and $\alpha$ is the electromagnetic fine structure constant. 

\begin{figure}[t]
\begin{minipage}{86mm}
\centering
\includegraphics[scale=0.64]{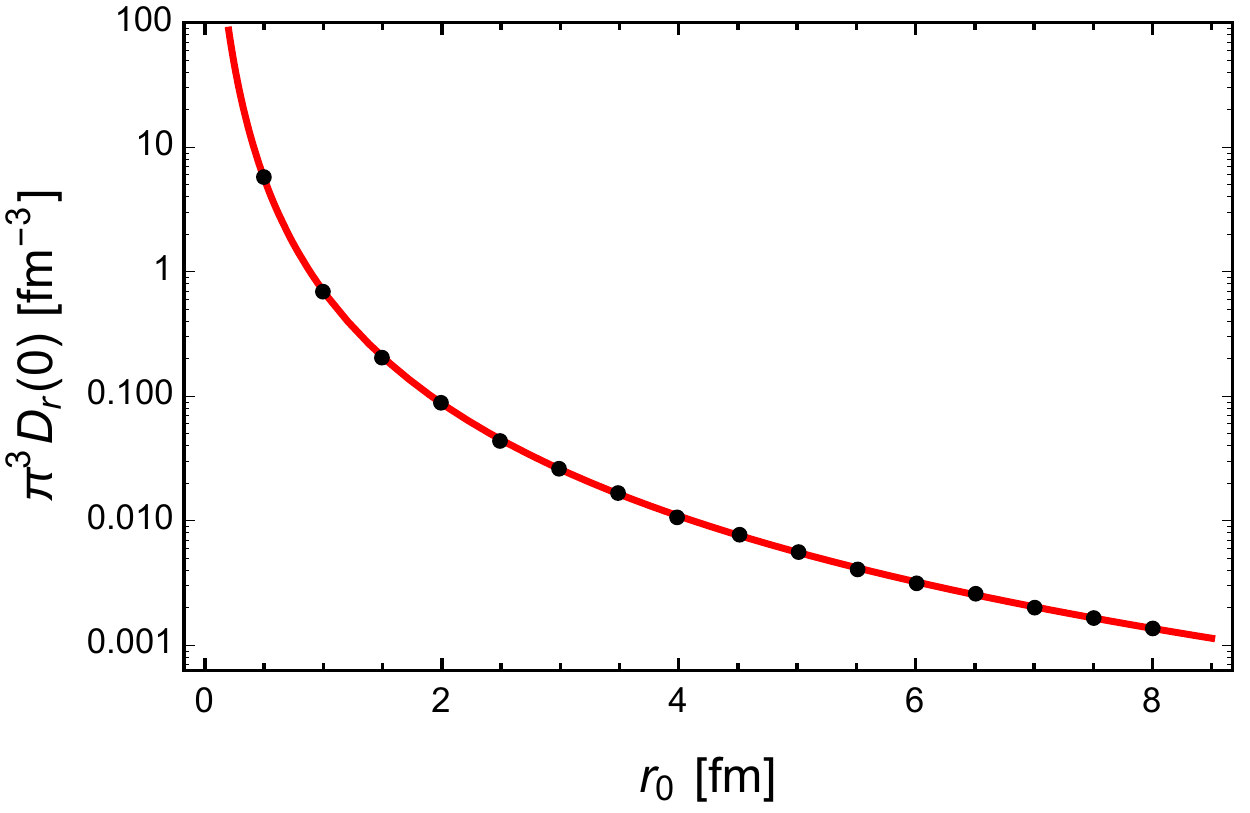}
\caption{The value of  $\pi^3 \, D_r (0)$ as a function of $r_0$ computed directly from the definition (\ref{Gauss-source-r}) (line) and from the sum rule (\ref{sum-improved-pions}) (dots).}
\label{fig-D0}
\end{minipage}
\hspace{2mm}
\begin{minipage}{86mm}
\centering
\vspace{-3mm}
\includegraphics[scale=0.69]{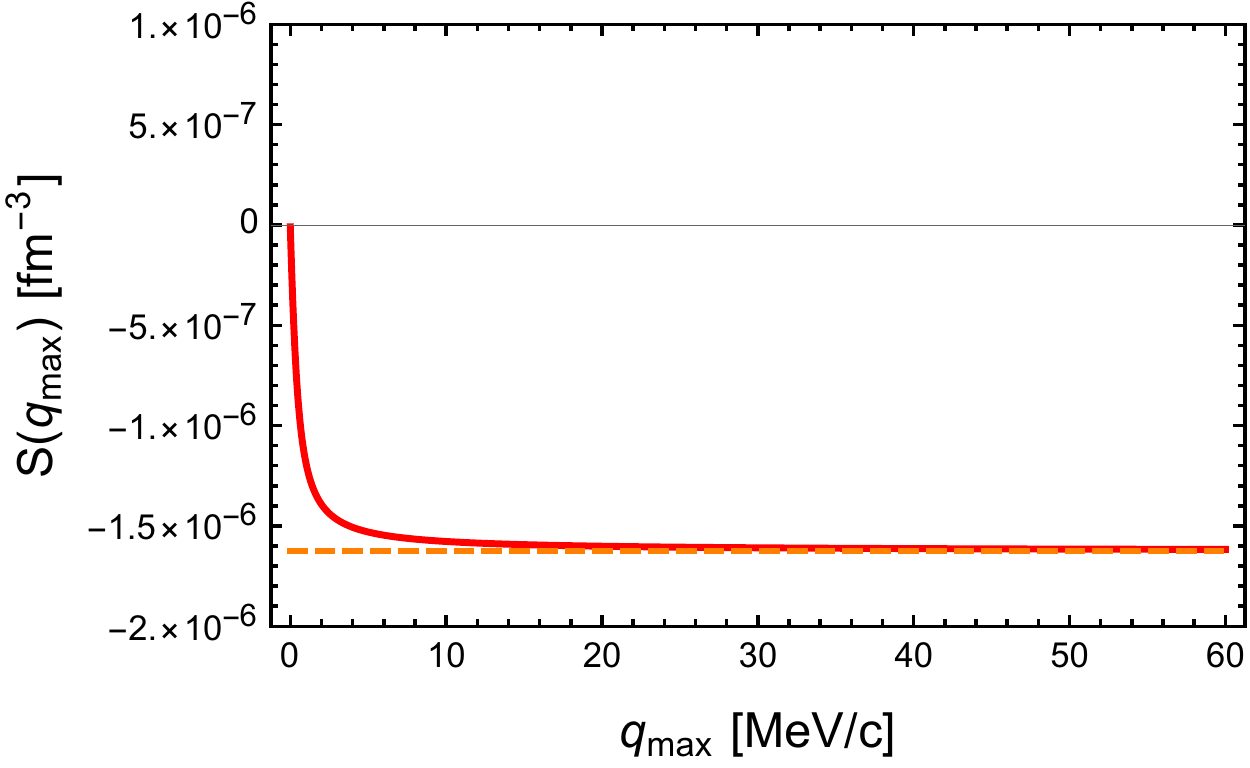}
\caption{The function  ${\cal S}(q_{\rm max})$ corresponding to Eq.~(\ref{sum-improved-pions-att-point-ren}) (solid) and the value of $- 8\pi^2 \zeta(3)/a_B^3$ (dashed).}
\label{fig-smax-ren}
\end{minipage}
\end{figure}

When the source function is point-like, the formation rate (\ref{form-rate}) of the  $\pi^+ \pi^-$  atom in a state $(n,m,l)$ is 
\be 
\label{form-rate-atom}
A_{nlm} =  (2\pi)^3 \vert \phi_{nlm}({\bf r}=0) \vert ^2 .
\ee
One should remember that the atomic wave function $ \phi_{nlm}({\bf r})$ behaves as $r^l$ when $r \to 0$ \cite{Schiff68} and consequently, the function is nonzero at ${\bf r}=0$ only for $l=0$. Since $\vert \phi_{n00}({\bf r}=0) \vert ^2 = 1/(\pi a_B^3 n^3)$, the expected sum rule (\ref{sum-improved-pions-att}) is
\be 
\label{sum-improved-pions-att-point}
4\pi \int_0^\infty dq \,q^2  \Big( G^+(q) +  G^-(q) - 2 \Big) = - \frac{8\pi^2}{a_B^3} \,\zeta(3) , 
\ee
where the zeta Riemann function $\zeta(3)$ is
\be
\zeta(3) \equiv  \sum_{n=1}^\infty \frac{1}{n^3} \approx 1.202 .
\ee

In spite of applying the regulator, the momentum integral in the l.h.s. of Eq.~(\ref{sum-improved-pions-att-point}) is still ultraviolet divergent. Since the Gamov factors $G^{\pm}(q)$ expanded in powers of $x \equiv 2\pi/(a_B q)$ are
\be
G^{\pm}(q) = 1 \mp \frac{x}{2}+\frac{x^2}{12} - \frac{x^4}{720}+\frac{x^6}{30240}+ {\cal O}(x^{8}) ,
\ee
the integrand in  Eq.~(\ref{sum-improved-pions-att-point}) is
\be 
\label{integrand-exp}
q^2  \Big( G^+(q) +  G^-(q) - 2 \Big)  = 
q^2 \bigg( \frac{1}{6} \Big(\frac{2\pi}{a_B q}\Big)^2 - \frac{1}{360} \Big(\frac{2\pi}{a_B q}\Big)^4 + {\cal O}(x^6)  \bigg) .
\ee
The first term in the series (\ref{integrand-exp}) is responsible for the linear ultraviolet divergence of the integral from Eq.~(\ref{sum-improved-pions-att-point}). Subtracting the term from the integrand, the sum rule (\ref{sum-improved-pions-att-point}) is modified to  
\be 
\label{sum-improved-pions-att-point-ren}
4\pi \int_0^\infty dq \,q^2  \Big( G^+(q) +  G^-(q) - 2 - \frac{2\pi^2}{3 a_B^2 q^2} \Big)  = -  \frac{8\pi^2}{a_B^3} \,\zeta(3) . 
\ee
In Fig.~\ref{fig-smax-ren} we show the function ${\cal S}(q_{\rm max})$ corresponding to the l.h.s. of Eq.~(\ref{sum-improved-pions-att-point-ren}) together with the value of $- 8\pi^2 \zeta(3)/a_B^3$. The figure clearly shows that the sum rule (\ref{sum-improved-pions-att-point-ren}) is satisfied for a sufficiently big $q_{\rm max}$. The subtraction method applied here suggests that a regulator may not be a physical correlation function.

\section{Neutron-proton correlation function}
\label{sec-np}

\begin{figure}[t]
\begin{minipage}{86mm}
\centering
\includegraphics[scale=0.67]{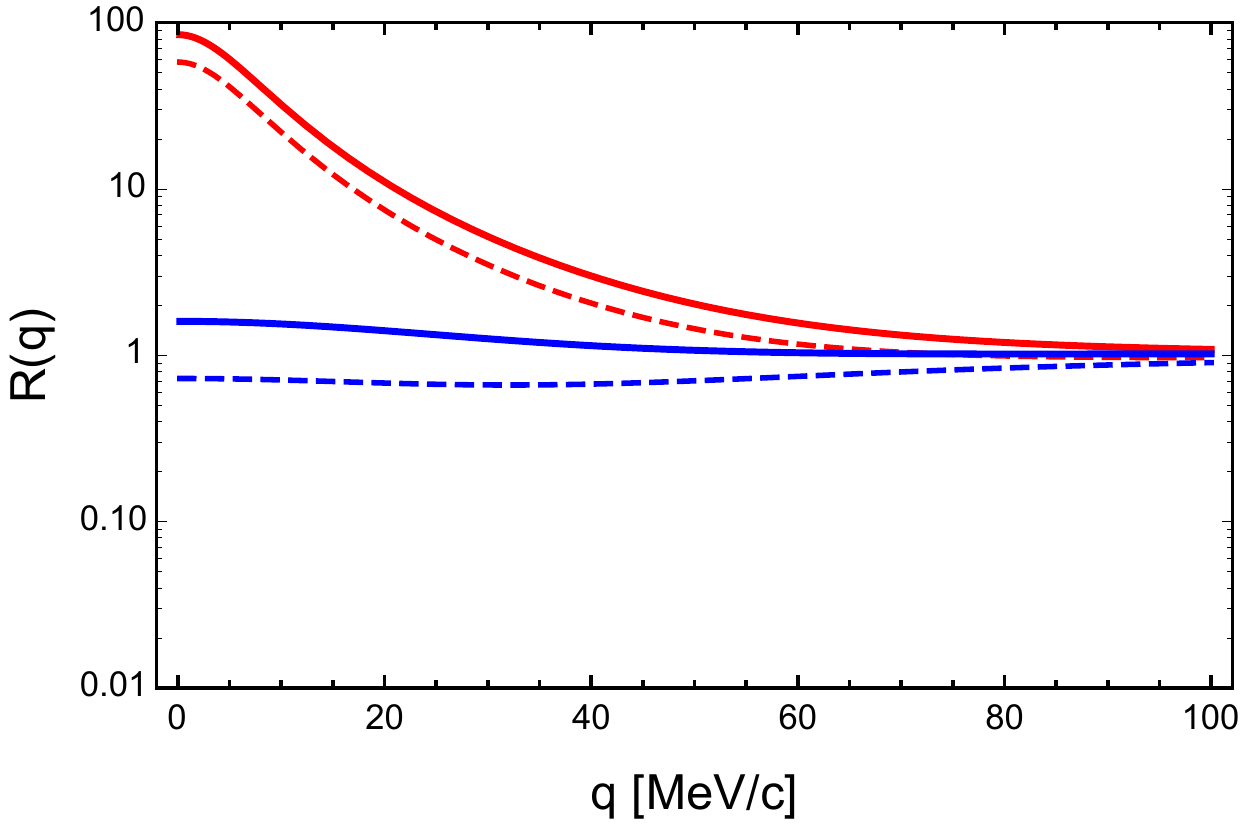}
\caption{The triplet (blue) and singlet (red) correlation functions of neutron and proton without (solid) and with (dashed) the correction factor (\ref{correction})  for $r_0 = 2$~fm.}
\label{fig-np-corr-2}
\end{minipage}
\hspace{3mm}
\begin{minipage}{86mm}
\centering
\includegraphics[scale=0.67]{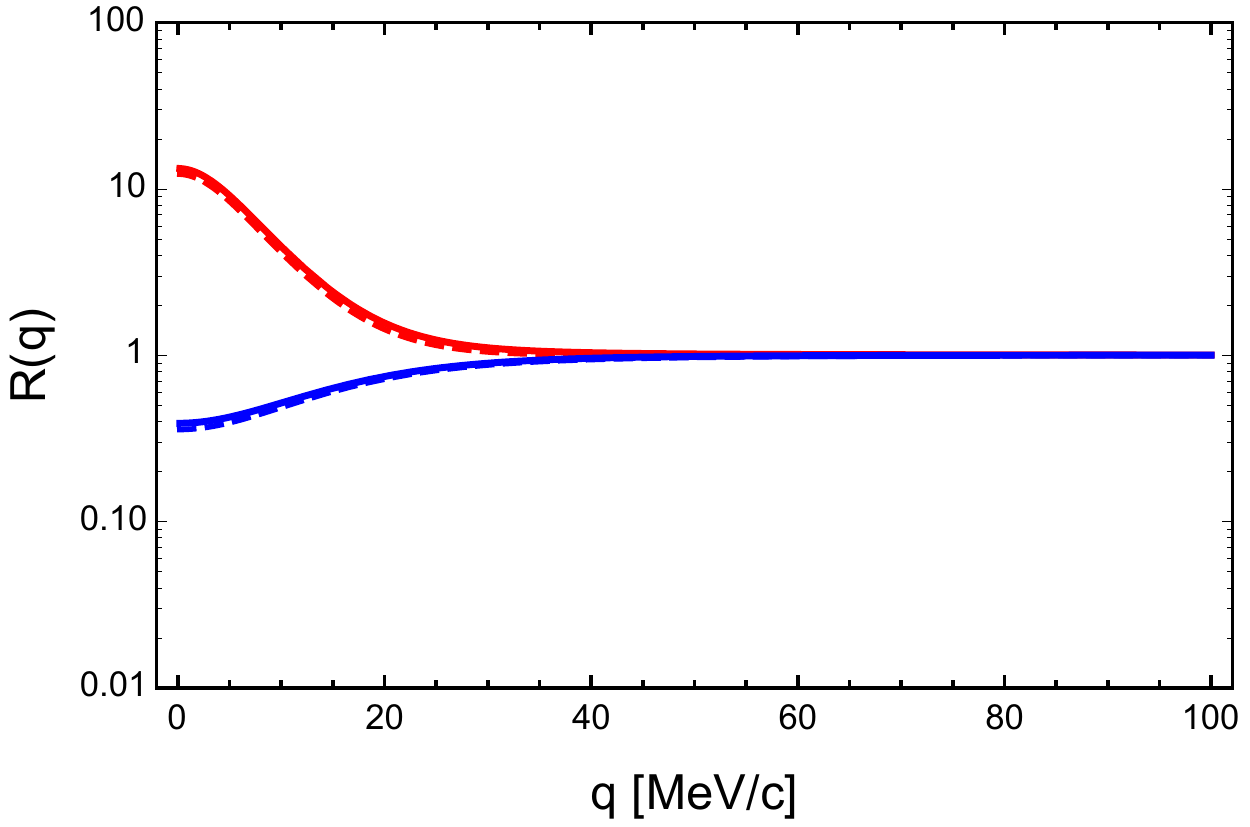}
\caption{The triplet (blue) and singlet (red) correlation functions of neutron and proton without (solid) and with (dashed) the correction factor (\ref{correction})  for $r_0 = 6$~fm.}
\label{fig-np-corr-6}
\end{minipage}
\hspace{2mm}
\end{figure}

Once we know that the sum rule  (\ref{sum-improved-pions}) works properly it can be used to check some model calculations of correlation functions. As an example we consider the neutron-proton correlation function obtained by means of the frequently used method proposed by Lednicky and Lyuboshits \cite{Lednicky:1981su}. 

A neutron-proton interaction depends on a spin state of the pair. A deuteron is a neutron-proton pair in a triplet state but there is no bound state in a singlet state. Therefore, we consider the correlation function of neutron and proton in a triplet state as ${\cal R}(q)$ and as the regulator $\widetilde{\cal R} (q)$  the correlation function of neutron and proton in a singlet state. 

According to the sum rule  (\ref{sum-improved}), the expected relation is 
\be 
\label{sum-improved-np}
4\pi \int_0^\infty dq \,q^2  \Big( {\cal R} (q) - \widetilde{\cal R} (q) \Big)
= -  A_D , 
\ee
where $A_D$ is the deuteron formation rate given by Eq.~(\ref{form-rate}). The rate is often computed with the Gaussian wave function of a deuteron 
\be
\label{Gauss}
|\phi_D ({\bf r}) |^2 = \frac{1}{(4 \pi r^2_D)^{3/2}} \, \exp \Big({-\frac{{\bf r}^2}{4r^2_D}} \Big),
\ee
where $r_D$ is the root-mean-square radius of a deuteron equal 2.0~fm \cite{Babenko:2008zz}. The formation rate then equals
\be
\label{A-Gauss}
A_D =  \frac{\pi^{3/2}}{(r_0^2 + r_D^2)^{3/2}} .
\ee 
A deuteron is much better described with the Hulth\' en wave function
\be 
\label{Hulthen} 
\phi _D({\bf r}) = \Big ( 
{\alpha \beta (\alpha + \beta) \over 2\pi (\alpha - \beta )^2} \Big )^{1/2} 
\;\; {e^{-\alpha r}-e^{-\beta r}  \over r } , 
\ee
where $\alpha = 0.23$ fm$^{-1}$ and $\beta = 1.61$ fm$^{-1}$ \cite{Hodgson71}, but the formation rate needs to be computed numerically. 

Assuming that the source radius is significantly larger than the n-p interaction range, the singlet $(s)$ or triplet $(t)$ wave function of the n-p pair in a scattering state is approximated by its asymptotic form
\be 
\label{scatter}
\phi _{np}^{s,t}({\bf r}) = e^{i{\bf q}{\bf r}}  + f^{s,t}(q)
{e^{iq r} \over r  } , 
\ee
where $f^{s,t}(q)$ is the scattering amplitude. It is chosen as
\be 
\label{ampli}
f^{s,t} (q)  = { -a^{s,t} \over 1 - {1 \over 2} d^{s,t}a^{s,t}q^2 
+ i q a^{s,t} } \;, 
\ee 
where $a^{s,t}$ and $d^{s,t}$ are the scattering lengths and the effective ranges. The parameters are $a^s = - 23.7$ fm, $d^s = 2.7$ fm and $a^t =  5.4$ fm, $d^t = 1.7$ fm \cite{McCarthy68}. The amplitude (\ref{ampli}) takes into account only the $s-$wave scattering. This is justified as long as only small relative velocities are considered.

\begin{figure}[t]
\begin{minipage}{86mm}
\centering
\includegraphics[scale=0.67]{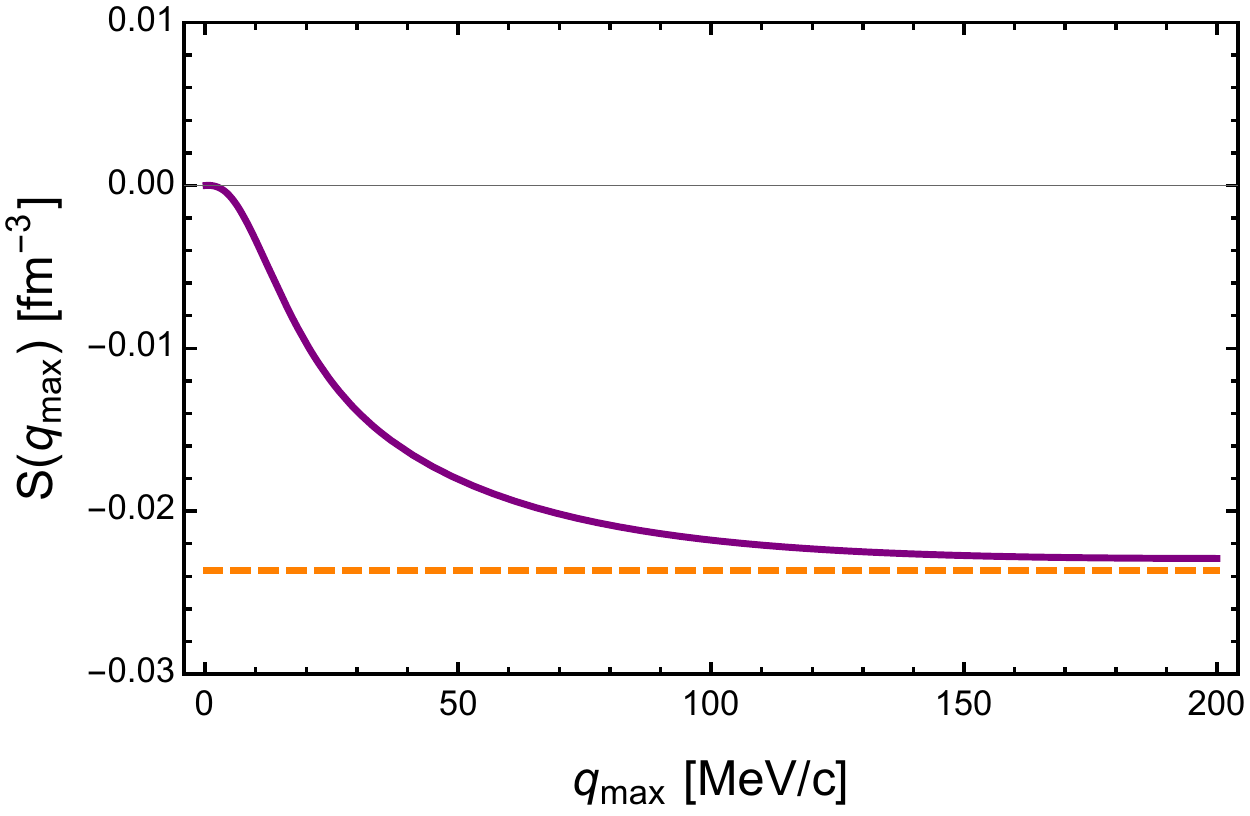}
\caption{The function ${\cal S}(q_{\rm max})$ corresponding to Eq.~(\ref{sum-improved-np}) (solid) and the value of $-A_D$ (dashed) for $r_0 = 6$~fm.}
\label{fig-smax-np}
\end{minipage}
\hspace{2mm}
\begin{minipage}{86mm}
\centering
\vspace{11mm}
\includegraphics[scale=0.68]{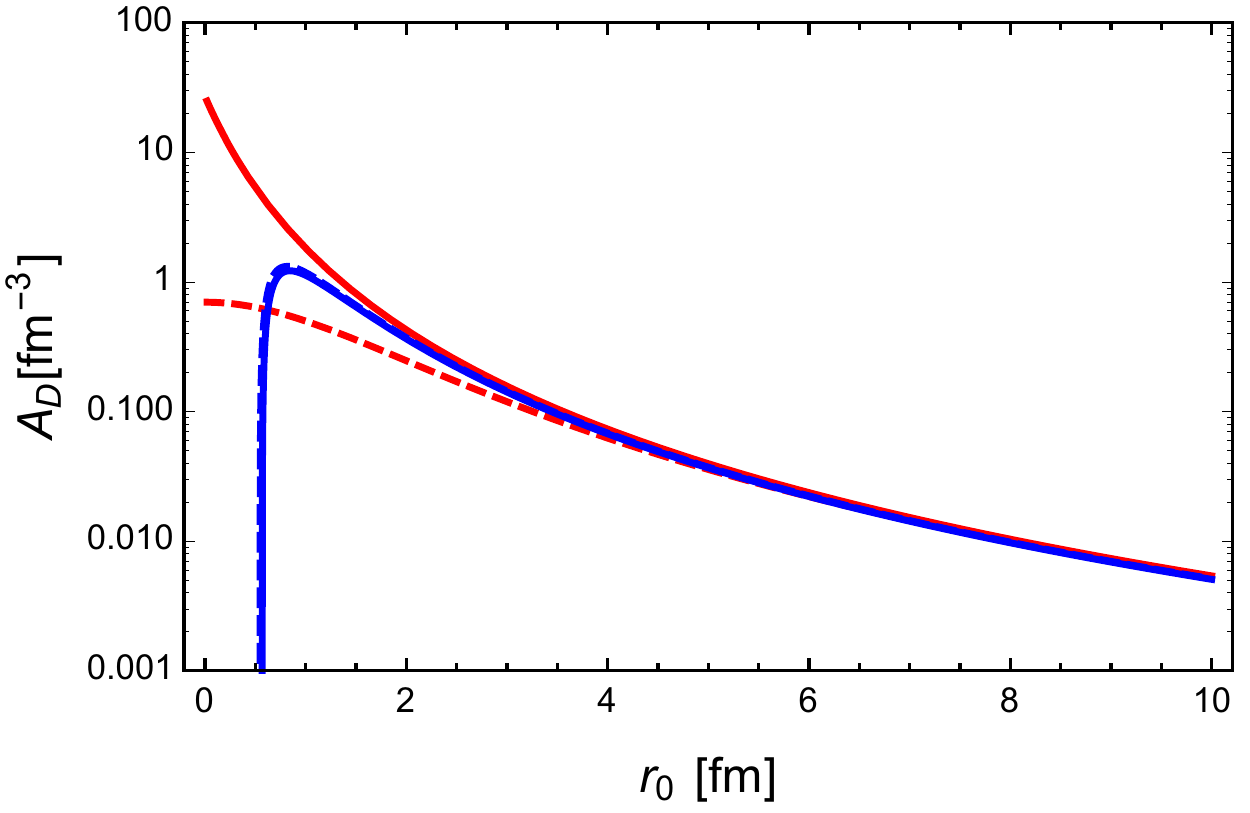}
\caption{The deuteron formation rate computed directly (red) using the Hulth\'en (solid) and Gaussian (dashed) wave functions and the rate obtained from the correlation function (blue) without (solid) and with (dashed) the correction factor (\ref{correction}).}
\label{fig-AD}
\end{minipage}
\end{figure}

The asymptotic wave function (\ref{scatter}) is used under the assumption that the source radius is significantly bigger than the interaction range. It was argued by Lednicky and Lyuboshits \cite{Lednicky:1981su} that the wave function (\ref{scatter}) can be still applied for smaller sources but the term of the correlation function proportional to $|f^{s,t}(q)|^2$ should be multiplied by the correction factor
\be
\label{correction}
1 - \frac{d^{s,t}}{2 \sqrt{\pi} \, r_0} , 
\ee
which is nowadays often used, see {\it e.g.} \cite{Acharya:2019ldv}. 

Substituting the wave function (\ref{scatter}) with the source function (\ref{Gauss-source-r}) into the formula (\ref{fun-corr-relative}), one gets the triplet and singlet correlation functions which are shown in Figs.~\ref{fig-np-corr-2} and \ref{fig-np-corr-6} for $r_0 =2$~fm and $r_0 =6$~fm. There are shown the functions without and with the correction factor (\ref{correction}) which indeed matters for the case of the smaller source. 

Fig.~\ref{fig-smax-np} presents the function ${\cal S}(q_{\rm max})$ defined by Eq.~(\ref{smax-improved}) and corresponding to the relation (\ref{sum-improved-np}). The n-p correlation functions are computed for $r_0 = 6$~fm without the correction factor (\ref{correction}). The value of $-A_D$ shown in Fig.~\ref{fig-smax-np} is obtained with the Hulth\' en wave function (\ref{Hulthen}). The function ${\cal S}(q_{\rm max})$ is seen to tend to $-A_D$ as $q_{\rm max}$ grows. Let us note here that an approximate formula, which relates the deuteron formation to the neutron-proton correlation, can be also found on a phenomenological ground \cite{Blum:2019suo}. 

In Fig.~\ref{fig-AD} we compare the deuteron formation rate computed directly from Eq.~(\ref{form-rate}) with the Gauss or Hulth\' en wave function to the rate obtained by means of the relation (\ref{sum-improved-np}) with and without the correction factor (\ref{correction}). One sees that the sum rule (\ref{sum-improved-np}) is fulfilled for $r_0$ bigger than 2 fm if the formation rate is computed with the Hulth\' en wave function. As expected, the Gaussian wave function is less accurate. 

Fig.~\ref{fig-np-corr-2} clearly shows that the correction factor (\ref{correction}) significantly modifies the correlation functions for smaller sources.  However, the factor does not help to satisfy the sum rule for smaller sources with $r_0 \le 1$~fm. The asymptotic form of the wave function (\ref{scatter}) is no longer a reliable approximation. The sum rule provides the quantitative criterion of applicability of the formula (\ref{scatter}).

\section{Summary and Conclusions}

The sum rule of femtoscopic correlation functions derived in \cite{Mrowczynski:1994rn} suffered from the ultraviolet divergence of momentum integral in physically interesting cases. We have resolved the problem by considering the sum rule not of a single correlation function but of the sum or difference of two appropriately chosen correlation functions. The ultraviolet divergence is then canceled out and the physical content of the sum rule remains similar. The improved sum rule has been shown to work properly for the exact Coulomb correlation functions in both cases of electrostatic repulsion and attraction. 

The improved sum rule can be used to test an accuracy and range of applicability of correlation functions computed in approximate models. We have considered the neutron-proton correlation function as an example. We showed that the asymptotic form of the wave function (\ref{scatter}), which is used to compute the correlation functions, is a reliable approximation for the particle source parameter $r_0$ bigger than 2 fm. The correction factor (\ref{correction}) does not much improve the situation. 

Nowadays when femtoscopic methods allow one to measure radii of particle sources and parameters of inter-particle interactions with a high precision, see \cite{Acharya:2018gyz} and \cite{Acharya:2019ldv}, respectively, the improved sum rule can be a useful tool to test an accuracy of theoretical models which are applied to compute correlation functions. 

\vspace{-3mm}

\acknowledgements

This work was partially supported by the National Science Centre, Poland under grant 2018/29/B/ST2/00646. The numerical calculations have been performed using {\it Mathematica}.


\end{document}